\documentclass{article}%
\usepackage{amsmath}
\usepackage{graphicx}%
\usepackage{amsfonts}%
\usepackage{amssymb}

\renewcommand{\P}{{\mathsf P}}

\renewcommand{\H}{{\mathcal H}}
\renewcommand{\O}{{\mathcal O}}
\newcommand{\B}{{\mathcal B}}
\renewcommand{\L}{{\mathcal L}}
\newcommand{\K}{{\mathcal K}}
\renewcommand{\a}{\alpha}
\newcommand{\R}{{\mathcal R}}
\renewcommand{\S}{\Sigma}

\newcommand{\unit}{{\mathbf 1}}

\newcommand{\id}{{\mathbf 1}}
\newcommand{\tr}{Tr}
\newcommand{\cp}{{\mathsf P}}

\begin{document}

\title{Status of the Fundamental Laws of Thermodynamics}

\author{Walid K. Abou Salem\footnote{Current address: Department of Mathematics, Univesity of Toronto, M5S 2E4 Toronto, Canada} and J\"urg Fr\"ohlich \\
Institute for Theoretical Physics \\
ETH Zurich \\
CH-8093 Switzerland \footnote{\it e-mail: walid@itp.phys.ethz.ch, juerg@itp.phys.ethz.ch}}

\date{ \; }

\maketitle

\begin{abstract}
We describe recent progress towards deriving the Fundamental Laws of thermodynamics (the $0^{th}, 1^{st},$ and $2^{nd}$ Law) from nonequilibrium quantum statistical mechanics in simple, yet physically relevant models. Along the way, we clarify some basic thermodynamic notions and discuss various {\it reversible} and {\it irreversible} thermodynamic processes from the point of view of quantum statistical mechanics. 
\end{abstract}


\section{Introduction}

Most systems in Nature can be viewed as interacting many-particle systems: Atoms and molecules in gases, fluids, superfluids and solids, electrons and ions in plasmas, the electron fluids in conductors and semi-conductors, nuclear
matter in neutron stars, etc. It is fascinating and
intriguing that certain aspects of {\it all} these systems, when close to thermal equilibrium, can be described by a few {\it general} and {\it universal} laws: the Fundamental Laws of thermodynamics. The purpose of thermodynamics is to describe average statistical properties of macroscopically extended systems of matter in states close to thermal equilibrium states, with small spatial and slow temporal variations. Typical macroscopic systems are formed of $10^{23}-10^{28}$ particles. Describing such systems microscopically, by solving the corresponding Hamilton equations or a Schr\"odinger equation, is a dauntingly difficult, in practice an impossible task. To circumvent this problem, one limits one's attention to describing emergent properties involving only a few observable macroscopic quantities, such as the volume $V$ of the system, its internal energy $U$, or its magnetization $M.$ These macroscopic quantities, which can be measured simultaneously, and, in principle, with arbitrary precision, are called extensive {\it thermodynamic observables}. The Laws of Thermodynamics give non-trivial relations between these quantities valid for arbitrary macroscopic systems. Thermodynamics is a highly successful physical theory that is self-contained. Nevertheless, it is interesting to attempt to derive its Fundamental Laws from an ab-initio microscopic description, in particular, from a quantum statistical description of many-particle systems. Conceptually, this is not only important to improve our understanding of thermodynamics, but it is also a consistency check of quantum statistical mechanics.

The program to derive the $0^{th}$, $1^{st}$, and $2^{nd}$ Law from kinetic theory and statistical mechanics
has been studied since the late $19^{th}$ Century, with
contributions by many distinguished scientists, including Maxwell, Boltzmann, Gibbs and Einstein. However, this program has not been completed so far. In this paper, we present some recent results summarizing our own attempts to
derive thermodynamics from quantum statistical
mechanics and to bring the program just described closer
to a satisfactory completion. We claim that, indeed, these laws can be derived in a mathematically rigorous manner from quantum statistical mechanics, provided one adopts a suitable notion of thermal reservoirs, assumes part of the $0^{th}$ law for such reservoirs, and limits the scope of the study to a class of idealized, yet physically relevant models. A more detailed presentation together with a discussion of Fourier's Law will appear in [A-SF4], (see also [BLR]).

Recent rigorous results concerning a partial derivation of the Fundamental Laws of thermodynamics from quantum statistical mechanics also represent progress towards understanding {\it irreversible} behaviour of macroscopic open systems and the emergence of {\it classical} regimes on the basis of more {\it fundamental}, {\it time-reversal invariant} microscopic laws, such as those of quantum mechanics. They are also a step in the direction of understanding a limiting regime in the description of many-particle systems in which the atomistic constitution of matter becomes irrelevant, corresponding to the limit as the Boltzmann constant, $k_B,$ tends to 0.

\section{Basic Concepts and Laws of Thermodynamics}

Before starting to discuss our derivation of the Laws of Thermodynamics from nonequilibrium quantum statistical mechanics, we recall some basic concepts and notions of thermodynamics.

One of the basic notions of thermodynamics is that of an {\it isolated system}, ie, of a time-translation invariant system without any contact to or interaction with its environment. For such a system, the measured values of extensive thermodynamic observables are {\it time-independent} (stationary). It is a fact of experience- and a standard assumption of thermodynamics- that the state of a {\it macroscopically large isolated} system approaches a state which, {\it locally}, is indistinguishable from a stationary {\it equilibrium state}, as time, $t,$ tends to $\infty.$ This {\it `` equilibrium postulate''} is the subtle part of the $0^{th}$ Law of Thermodynamics. In our analysis, this part of the $0^{th}$ Law is used to justify our {\it assumption} that the state of an isolated, macroscopically large (infinite) heat bath always approaches a {\it thermal equilibrium state}, as $t\rightarrow\infty.$ This assumption will {\it not} be fully, but only {\it partially} proven for infinitely extended, dispersive heat baths. Infinitely extended, dispersive systems are called {\it open systems}.

Let $N$ be the number of elements in a complete family of independent extensive thermodynamic observables of a system $\S.$ These observables are conserved and can be measured simultaneously and with arbitrary precision. Their measured values specify a point $X\in \Gamma^\S$, where $\Gamma^\S$ is a convex subset of ${\mathbf R}^N.$ A thermodynamic observable is a real- valued function on $\Gamma^\S.$ Every point $X\in \Gamma^\S$ corresponds to a unique equilibrium state of $\S$. 

One may couple two thermodynamic systems, $1$ and $2$, through local interactions. When these interactions vanish, the state space of the coupled system is the Cartesian product, $\Gamma^{1}\times \Gamma^{2}.$ When one introduces  interactions between $1$ and $2,$ some symmetries of $1$ and $2$ can be broken, and the corresponding generators are not conserved quantities, anymore. The new family of extensive thermodynamic observables and the space of equilibrium states, $\Gamma^{1\vee 2},$ of the coupled system depend on the type of interactions between $1$ and $2,$ in particular on the symmetries preserved by the interaction. 


Next, we discuss the notion of a {\it thermodynamic process}, which plays a central role in thermodynamics. Let $(X_1, X_2)\in \Gamma^1\times\Gamma^2$ correspond to an initial product equilibrium state of $1\vee 2$ at some time $t_0,$ before the two systems are coupled. Suppose that an interaction between $1$ and $2$ is turned on at time $t_0.$ One is then interested in predicting the state of the coupled system at time $t_0+T$, as $T\rightarrow\infty.$ Let $\gamma(t)$ be the macrostate of the coupled system.


If one system is finite, and the other one is macroscopically large, the {\it ``equilibrium postulate''} says that $\gamma(t)$ converges to an equilibrium state $X_{12}\in \Gamma^{1\vee 2}$ of the coupled system, as $t\rightarrow\infty.$ In thermodynamics, the map
\begin{equation*}
\Gamma^1\times\Gamma^2 \ni (X_1,X_2)\mapsto X_{12}\in \Gamma^{12}
\end{equation*}
is only predictable if the interactions between 1 and 2 are specified, or, put differently, if it is specified which constraints on $1\vee 2$ are eliminated through the coupling. 

If the reverse process,
\begin{equation*}
X_{12}\mapsto (X_1,X_2)
\end{equation*}
{\it cannot} be realized {\it without} coupling the system $1\vee 2$ to further macroscopically large systems, we say that the process $(X_1,X_2)\mapsto X_{12}$ is {\it irreversible}.

A thermodynamic process $\{ \gamma (t)\}_{t_0\le t<\infty}$ of a system $\S$ is {\it reversible} iff $\gamma(t)=X(t)\in \Gamma^\S$ is an equilibrium state of $\S,$ for all $t\in [t_0,\infty).$ Of course, this is an {\it idealized notion}. In practice, $\gamma(t)$ can only be very close to, but not identical to, an equilibrium state, for $t_0<t<\infty.$ We set $X_i:=\gamma(t_0)$ (initial state of $\S$), and $X_f:= \lim_{t\rightarrow\infty}\gamma(t)$ (final state of $\S$). In thermodynamics, one is only interested in predicting $X_f,$ assuming one knows the nature of the process and its initial state $X_i.$

As mentioned above, {\it extensive} thermodynamic observables of a system $\S$ correspond to {\it conserved quantities} (conservation laws) of $\S.$ These conserved quantities are generators of symmetries of $\S.$ The moduli space, $\Gamma^\S,$ of equilibrium states of $\S,$ is the {\it convex closure of the joint spectrum} of a maximal family of independent conserved quantities that can be measured {\it simultaneously}. To each symmetry of $\S$ that remains {\it unbroken} in a thermodynamic process of $\S,$ there corresponds an extensive thermodynamic observable whose value remains constant in time. One can thus classify thermodynamic processes according to the symmetries they leave unbroken. {\it ``Eliminating a constraint''} amounts to turning on interactions between subsystems of $\S$ that break one or several of the original symmetries of $\S,$ (but may leave other symmetries unbroken).

A {\it thermal contact} (diathermal wall) between a thermodynamic system $\S$ and a thermal reservoir $\R$ is an interaction which leaves all symmetries of $\S$ unbroken {\it except} for {\it time-translation invariance}. It leaves all the thermodynamic observables of $\S$ unchanged except for its {\it energy}. Similarly, one can define a thermal contact between two thermodynamic systems $\S_1$ and $\S_2$ as an interaction which preserves all the symmetries of $\S_1$ and $\S_2,$ except for time-translation invariance: It leaves all the thermodynamic observables of $\S_1$ and $\S_2$ invariant except for their energies. 

\subsection{Laws of Thermodynamics}

In this subsection, we recall the fundamental Laws of Thermodynamics ($0^{th},1^{st},$ and $2^{nd}$ Law), which form the axiomatic basis of thermodynamics.\footnote{We will not discuss the Third Law of thermodynamics.} We are interested in physical properties of a thermodynamic system $\S$ that can be encoded in a finite number, $N,$ of independent extensive thermodynamic observables, $\xi_1,\cdots, \xi_N.$

\bigskip
\noindent{\bf The $0^{th}$ Law}

\noindent{\it There are several parts to the $0^{th}$ Law.
\begin{itemize}

\item[(i)] There exist, for all practical purposes, infinitely large thermodynamic systems that approach thermal  equilibrium when isolated from their environment. Such systems are called (thermal) reservoirs or heat baths. Two thermal reservoirs, $\R_1$ and $\R_2,$ are said to be equivalent ($\R_1\sim \R_2$) {\it iff} no energy flows between $\R_1$ and $\R_2$ when a diathermal contact is established between them. We then say that the two reservoirs $\R_1$ and $\R_2$ have the same temperature. Furthermore, given three thermal reservoirs, $\R_1,\R_2$ and $\R_3,$ such that $\R_1\sim \R_2$ and $\R_2\sim \R_3,$ then $\R_1\sim \R_3$, ie, the equivalence of heat baths is transitive.

\item[(ii)]  When one brings a finite thermodynamic system $\S$ in thermal contact with a reservoir $\R$ and waits for an (infinitely) long time the state of the coupled system will asymptotically converge to an equilibrium state at the temperature of the reservoir.

\item[(iii)] Moreover, if one turns off the contact between $\S$ and $\R$ quasi-statically (adiabatically) the final state of $\S$ is the equilibrium state at the temperature of the reservoir, while the final state of the reservoir is identical to its initial equilibrium state.

\end{itemize}}

\bigskip

In our derivation of the Laws of Thermodynamics from quantum statistical mechanics, we assume some portion of part (i) of the $0^{th}$ Law of Thermodynamics, while we are able to prove parts (ii) and (iii) for idealized, yet physically relevant model systems. The difficult portion of part (i) represents an open problem not unrelated to the one of understanding the dynamics of macroscopic systems with translationally invariant many-body interactions.

\bigskip
\noindent{\bf The $1^{st}$ Law}

\noindent{\it For each finite thermodynamic system $\S$, there exists a thermodynamic observable $U,$ the internal energy, which has a definite value in each state of $\S$; ($U$ is defined uniquely, up to an additive constant). For a thermodynamic process $\gamma$ in which one brings $\S$ in contact with a thermal reservoir $\R$, the total amount of  heat energy $\Delta Q(\gamma),$ exchanged between $\R$ and $\S$ in the course of the process $\gamma$, is a well-defined quantity which depends not only on the initial point, $X_i=\partial_i\gamma,$ and the final point, $X_f=\partial_f\gamma,$ of $\gamma,$ but on the whole trajectory $\gamma$. \footnote{If $\Delta Q>0$, heat energy flows from $\R$ to $\S,$ and if $\Delta Q<0$ heat flows from $\S$ to $\R.$} The difference
\begin{equation*}
\Delta A(\gamma):= U(X_f)-U(X_i)-\Delta Q(\gamma)\; ,
\end{equation*}
is the {\it work} done on $\S.$}
\bigskip

Before stating the $2^{nd}$ Law, we need to introduce the notion of a heat engine. A {\it heat engine} is a finite thermodynamic system that is driven {\it periodically in time} and that is brought in contact with at least two inequivalent thermal reservoirs and with its environment. After each cycle (or period), the system returns to its initial state, ie, $\partial_i\gamma=\partial_f\gamma.$ Let $\Delta Q(\gamma)$ be the total heat energy exchanged between the heat engine and the thermal reservoirs in one cycle. Since the internal energy of the heat engine is the same at the beginning and at the end of each cycle, the $1^{st}$ Law says that $\Delta Q(\gamma)$ is fully converted into work done by the heat engine on its environment.

Usually, one introduces the following (scaling) postulate on heat engines: The size of a heat engine can be enlarged or reduced by a scale factor $\lambda >0.$ (Here, a continuum theory of matter is implicitly assumed.) Consider a heat engine $\S$ with a moduli space of equilibrium states $\Gamma^\S.$ Then
\begin{equation*}
\Gamma^{\S^\lambda} := \{ X\in {\mathbf R}^N : \lambda^{-1} X\in \Gamma^\S \} \; 
\end{equation*}
is the moduli space of equilibrium states of the heat engine $\S^\lambda.$ To a cycle $\gamma$ of $\S,$ there corresponds a cycle $\gamma^\lambda$ of $\S^\lambda$ such that
\begin{equation*}
U(\lambda X)=\lambda U(X) , \; \Delta Q(\gamma^\lambda)=\lambda \Delta Q(\gamma) \; .
\end{equation*}

We are now in a position to state one formulation of the Second Law of Thermodynamics due to Thomson and Planck.

\bigskip
\noindent{\bf The $2^{nd}$ Law}

\noindent{\it There does not exist any heat engine that does nothing but absorb heat energy from one single reservoir and convert it into work.}

\bigskip

Consider a heat engine $\S$ connected to two thermal reservoirs, $\R_1$ and $\R_2,$ with the property that, in one cycle $\gamma$, it gains an amount $\Delta Q_1$ of heat energy from $\R_1$ and it releases an amount $\Delta Q_2$ of heat energy to $\R_2$. The heat engine performs work if $\Delta Q_1-|\Delta Q_2|=\Delta Q_1+\Delta Q_2>0.$ In this case, the thermal reservoir $\R_1$ is called the heating, while $\R_2$ is called the refrigerator.

It follows from the above formulation of the Second Law of Thermodynamics that if there exists a heat engine that uses $\R_1$ as its heating and $\R_2$ as its refrigerator, then there does not exist any heat engine that uses $\R_2$ as its heating and $\R_1$ as its refrigerator. This fact can be used to define an {\it empirical} temperature $\Theta :$ the temperature $\Theta_1$ of $\R_1$ is higher than the temperature $\Theta_2$ of $\R_2$ if there exists a heat engine $\S$ that uses $\R_1$ as its heating and $\R_2$ as its refrigerator.

A heat engine is said to be {\it reversible} (or a {\it Carnot machine}) if, in a time-reversed cycle, it can work as a {\it heat pump}: During a cycle $\gamma^-$, it takes an amount $\Delta Q_2$ of heat energy from $\R_2$ and releases an amount $\Delta Q_1$ of heat energy to $\R_1.$ The environment must supply an amount $\Delta A=\Delta Q_1-|\Delta Q_2|$ of work per cycle. Reversible heat engines are idealizations of realistic engines.

We define the {\it degree of efficiency} of a heat engine $\S$ as the ratio of the work done per cycle and the heat energy it gains from the heating in one cycle, ie,
\begin{equation*}
\eta^\S := \frac{\Delta A}{\Delta Q_1} =\frac{\Delta Q_1 + \Delta Q_2}{\Delta Q_1} = 1 + \frac{\Delta Q_2}{\Delta Q_1} \; .
\end{equation*}
It follows from the Second Law of thermodynamics that among all heat engines with the same heating and refrigerator, the reversible engines have the highest degree of efficiency, $\eta^{rev}.$ One can use this fact to define an {\it absolute temperature} $T$ of a thermal reservoir $\R$ by setting
\begin{equation*}
\eta^{rev}=\frac{T_1-T_2}{T_1}\; ,
\end{equation*}
for an arbitrary pair of heating and refrigerator.
The fact that $\eta^\S\le \eta^{rev}$ implies that
\begin{equation*}
\frac{\Delta Q_1}{T_1} + \frac{\Delta Q_2}{T_2}\le 0 \; ,
\end{equation*}
with equality when $\gamma$ is reversible.

This result can be generalized to a situation where $\S$ is connected to $n$ thermal reservoirs, $\R_1,\cdots ,\R_n,$ with temperatures $T_1>\cdots >T_n.$ Then
\begin{equation*}
\sum_{i=1}^n \frac{\Delta Q_i}{T_i}\le 0 \; ,
\end{equation*}
with equality if the cyclic process is {\it reversible}. Taking the limit $n\rightarrow\infty$ yields 
\begin{equation*}
\oint_\gamma \frac{\delta Q}{T} \le 0 \; ,
\end{equation*}
with equality if $\gamma$ is reversible.

Consider a reversible cyclic process, $\gamma\subset \Gamma^\S,$ of $\S,$ and parametrize its trajectory in $\Gamma^\S$ by time $\tau\in [t_0,\infty).$ We assume that
\begin{equation*}
\dot{\gamma}(\tau):= \lim_{h\searrow 0}\frac{1}{h} [\gamma (\tau+h)-\gamma(\tau )] \in T_{\gamma (\tau)} \Gamma^\S\subset {\mathbf R}^N \; 
\end{equation*}
exists, for all $\tau\in [t_0,\infty ).$

Denote by $\gamma_t$ the subprocess $\{ \gamma (\tau) \}_{t_0\le \tau \le t}$ from $X_i:=\gamma (t_0)$ to $\gamma (t)\in \Gamma^\S.$ From the $1^{st}$ Law of Thermodynamics, we infer that $\Delta Q(\gamma_t)$ is a well-defined quantity. For $h>0,$
\begin{equation*}
\Delta Q(\gamma_{t+h})-\Delta Q(\gamma_t)=h \cdot K(t)+O(h^2) \; ,
\end{equation*}
where (we assume) $K(t)$ is continuous in $t.$ For every point $X\in\Gamma^\S$ and each vector $Z\in {\mathbf R}^N,$ there exists a subprocess $\gamma_t$ of a reversible cyclic process $\gamma$ of $\S$ such that
\begin{equation*}
\gamma(t)=X \; ; \dot{\gamma}(t) =  Z \; .
\end{equation*}
One can use the functional $\Delta Q(\cdot)$ defined on the set of reversible processes of $\S$ to define a {\it 1-form} $\delta Q(\gamma (t))$ with the property that\footnote{These arguments need to be made mathematically accurate. For some details and references, see for example [LY].}
\begin{equation*}
\dot{\gamma}(t)\cdot \delta Q(\gamma(t)) = \lim_{h\searrow 0} \frac{1}{h}(\Delta Q(\gamma_{t+h})-\Delta Q(\gamma_t))=K(t) \; .
\end{equation*} 
The internal energy $U$ of $\S$ is a state function, hence a function on $\Gamma^\S.$ Denote by $dU$ the 1-form over $\Gamma^\S$ given by the gradient of $U.$ We define the work 1-form by
\begin{equation*}
\delta A := dU -\delta Q \; .
\end{equation*}
Let $X_1,\cdots, X_N$ be coordinates on $\Gamma^\S. $ Then one can write
\begin{equation*}
\delta A =\sum_{i=1}^N a_i(X) dX_i \; ,
\end{equation*}
where $a_i(X), i=1,\cdots, N,$ are called the work coefficients. They are {\it intensive quantities}, meaning that under rescaling, $a_i(\lambda X)=a_i(X), i=1,\cdots, N.$\footnote{Quantities $\xi$ with the property that under rescaling $\xi(\lambda X)=\lambda \xi(X), \lambda > 0,$ are called {\it extensive}, e.g., the internal energy $U$ or the volume, while quantities with the property that $\xi(\lambda X)=\xi(X)$ are called {\it intensive}, e.g., the temperature $T,$ and the work coefficients.}

Using the fact that
\begin{equation*}
\oint_{\gamma^{rev}} \frac{\delta Q}{T}=0 , \; \forall \; \gamma^{rev}\subset \Gamma^\S,
\end{equation*}
and the convexity of $\Gamma^\S,$ one can define a state function $S$, the {\it entropy}, on $\Gamma^\S$ such that
\begin{equation*}
dS=\frac{\delta Q}{T} \; .
\end{equation*}
Then
\begin{equation*}
dU= TdS + \delta A \; 
\end{equation*}
holds for reversible changes of state.

Consider an adiabatic process $\gamma: X_i\rightarrow X_f$ of an {\it isolated} system $\S,$ such that $X_{i,f}\in \Gamma^\S.$ It follows from the definition of entropy and the fact that $\oint_{\overline{\gamma}} \frac{\delta Q}{T}\le 0,$ for a cyclic process $\overline{\gamma}\supset \gamma,$ that
\begin{equation*}
S(X_f)\ge S(X_i) \; .
\end{equation*}
Using the scaling postulate and the connectedness and convexity of $\Gamma^\S,$ one can show that the entropy $S$ is concave: For $\lambda\in(0,1),$
\begin{equation*}
S(\lambda X_1 + (1-\lambda) X_2) \ge \lambda S(X_1) + (1-\lambda) S(X_2) \; .
\end{equation*}

There are further equivalent formulations of the Second Law of Thermodynamics:
{\it
\begin{itemize}
\item[(i)]{\it Clausius} (1854): Suppose two reservoirs, $\R_1$ and $\R_2,$ are connected diathermally. If heat flows between them then it can only flow in one direction.

\item[(ii)]{\it Carnot} (1824): For a heat engine $\S,$ $\eta^\S\le\eta^{rev}.$

\item[(iii)]{\it Caratheodory}: In an arbitrarily small neighborhood of each equilibrium state, $X,$ of an isolated system $\S,$ there are equilibrium states $X'$ of $\S$ that are not accessible from $X$ via reversible and adiabatic processes. \footnote{For a mathematically rigorous discussion, see for example [Boy], and also [LY].}

\end{itemize}}

It follows that, during an adiabatic process of an isolated system, the entropy can only increase (maximum principle for the entropy).

In the following sections, we will show how Clausius' and Carnot's formulation of the Second Law of Thermodynamics can be derived from quantum statistical mechanics in simple systems.


\section{Quantum mechanical description of thermodynamic systems, heat baths, and thermodynamic processes}


We start by clarifying the concept of a {\it thermodynamic system} $\S$ from the point of view of quantum statistical mechanics. A thermodynamic system is a system of quantum-mechanical matter confined to a compact region of space. Physical properties of the system $\S$ are encoded in bounded operators acting on a separable Hilbert space, $\H^\S,$ of pure state vectors. These operators generate some subalgebra, $\O^\S,$ of $\B(\H^\S)$, where $\B(\H^\S)$ is
the algebra of bounded operators on $\H^\S.$ The algebra $\O^\S$ is called the {\it kinematical algebra} of $\S$. The pure states of $\S$ are unit rays in $\H^\S$, and its mixed states are described by density matrices $\P$, which are positive trace-class operators such that $Tr (\P )=1$. The dynamics of $\S$ is generated
by a family of semi-bounded, self-adjoint operators $\{H^\S(t)\}_{t\in {\mathbf R}}$ acting on $\H^\S,$ the {\it Hamiltonians}. Under natural hypotheses, these operators determine a unitary propagator, $V^\S(t,s),$ describing the time evolution of a state of $\S$ at time $s$ to a corresponding state at time $t.$ In the Heisenberg picture, the time evolution of an operator $A\in \O^\S$ is given by
\begin{equation}
\a^{t,s}_\S (A) = V^\S(s,t) A V^\S(t,s) \; , \label{dym1}
\end{equation}
and we assume that $\a^{t,s}_\S(A)\in\O^\S,$ for every $A\in O^\S$.
 Since a density matrix $\P$ describing a mixed state of $\S$ is a positive, trace-class operator on $\H^\S,$ it has a square-root $\kappa=\P^{\frac{1}{2}}$ belonging to $\L^2(\H^\S
)=:\K^\S$, the two-sided ideal of Hilbert-Schmidt operators in $\B(\H^\S
),$ which is isomorphic to $\H^\S\otimes\H^\S$. Then 
\begin{equation}
Tr (\P A) = Tr (\kappa^* A \kappa):= \langle \kappa, A\kappa\rangle .
\end{equation}
In the Schr\"odinger picture, the time-evolution of a state
$\kappa_s\in \K^\S$ from time $s$ to time $t$ is given by is 
$$\kappa_t=U_\S (t,s)\kappa_s:= V^\S (t,s)\kappa_s V^\S (s,t).$$
Then
$$\langle \kappa_t, A \kappa_t \rangle = \langle \kappa_s, \alpha^{t,s}_\S (A)\kappa_s \rangle ,\forall A\in \O^\S.$$
The propagator $U_\S(t,s),$ on $\K^\S$ is generated by a family of (usually time-dependent) {\it Liouvilleans} $\{\L^\S(t)\},$ with $\L^\S(t)=ad_{H^\S(t)}.$ It satisfies the equation
\begin{equation}
\partial_t U_\S(t,s)=-i \L^\S(t) U_\S(t,s) ,
\end{equation}
with $U_\S(s,s)={\mathbf 1},\forall s.$\footnote{We work in units where $\hbar=1.$} Since $\K^\S$ is a Hilbert space, one may study the spectra of the Liouvilleans $\L^\S(t)$ using available methods of spectral theory.

The formulation outlined here has a natural incarnation in the thermodynamic limit of systems in thermal equilibrium; see [HHW,BFS].

According to the Gibbs Ansatz, the equilibrium state of $\S$ at
inverse temperature $\beta>0$ is described, in the canonical ensemble, by the density matrix
\begin{equation}
{\mathsf P}^\S_\beta := \frac{e^{-\beta H^\S}}{Z^\S_\beta} \;
,
\end{equation}
where $Z^\S_\beta := Tr (e^{-\beta H^\S})$ is a normalization
factor. The expectation value of an operator $A\in \O^\S$ in this equilibrium state 
is given by
\begin{equation}
\omega^\S_\beta (A):= Tr ({\mathsf P}^\S_\beta A) \; . \label{equil1}
\end{equation}
Note that if $H^\S(t)=H^\S$ is independent of time then $\omega_\beta^\S$ is time-translation invariant and satisfies the Kubo- Martin- Schwinger (KMS) condition, which will be recalled later.

We distinguish between two types of thermodynamic systems, those with a finite-dimensional Hilbert space (mesoscopic systems, such as an {\it impurity spin} or a {\it quantum dot}), and macroscopic systems, which have a countably infinite-dimensional Hilbert space. Understanding how the state of an isolated macroscopic system converges to a state that, locally, is indistinguishable from an equilibrium state, as time tends to infinity, is usually a challenging open problem. Macroscopic systems are defined in terms of {\it families} of thermodynamic systems, $\S_i,$ confined to regions $\Lambda_i,$ with $\Lambda_i\nearrow {\mathbf R}^3,$ with the property that $\{ \S_i\}$ is thermodynamically {\it stable}. \footnote{A brief remark about thermodynamic stability is in order at this point; (for further discussion, see for example [Ru1]). For a thermodynamic system $\S$ given by the disjoint union of elements of a family of thermodynamic systems, $\{\S_i\},$  $\S=\bigvee_i \S_i,$ the Hilbert space of $\S$ is given by $\H^\S=\otimes_i \H^{\S_i},$ and the kinematical algebra of $\S$ is given by $\O^\S=\overline{\otimes_i \O^{\S_i}}.$ The Hamiltonian of $\S$ is $H_0^\S=\sum_i H_0^{\S_i}+surface \; terms,$ with the property that $H_0^{\S_i}\approx H_0^{\S_j}$ if $\S_i$ is the spatial translate of $\S_j.$ We say that $\S$ is thermodynamically stable if $Tr (e^{-\beta H_0^{\S}})\le e^{C_\beta vol(\Lambda^\S)},$ as $\Lambda^\S\nearrow {\mathbf R}^3,$ $\forall \beta.$}


A {\it heat bath} or {\it reservoir} $\R$ is the limit of a sequence
of thermodynamic systems confined to compact regions of physical
space ${\mathbf R}^3,$ $\{ \Lambda_i \}_{i=1}^\infty$, such that
$\Lambda_i\subseteq \Lambda_j\subset {\mathbf R}^3,$ for $i<j,$ and
$\lim_{i\rightarrow\infty}\Lambda_i = {\mathbf R}^3,$ or a {\it half-space} ${\mathbf R}^3_\pm.$ The Hamiltonians $H^{\Lambda_i}$ are assumed to be time-independent. Denote by
$\O^{\Lambda_i}$ the kinematical algebra of the system confined to $\Lambda_i.$ We assume that $\O^{\Lambda_i}\subseteq \O^{\Lambda_j}$ if
$i<j$. The kinematical algebra of the heat bath $\R$ is $\O^\R:= \overline{\bigvee _{i\in {\mathbf
N}}\O^{\Lambda_i}}$, where $\overline{( \cdot )}$ denotes the norm
closure.

We make the following assumptions, which need to be verified in
specific physical models, regarding the existence of the time evolution and
equilibrium states in the thermodynamic limit; (see [BR,Ru1] for models where the following assumptions are verified). Let $\O^\infty :=
\bigvee_{i\in {\mathbf N}}\O^{\Lambda_i}$.

\begin{itemize}

\item[(A1)] {\it Existence of dynamics.}
We assume that
\begin{equation}
n-\lim_{i} \a_{\Lambda_i}^t (A) = : \a_\R^t (A) \; , \label{dyn2}
\end{equation}
exists for all $A\in \O^\infty$, $t\in {\mathbf R}$, and $\{ \a_\R^t
\}_{t\in {\mathbf R}}$ is a one-parameter group of $*$-
automorphisms of $\O^\R$. (Note that $\a_\R^t$ need not be norm
continuous, as in the case of bosonic reservoirs, where it is only
$\sigma$-weakly continuous, [BR].)

\item[(A2)]{\it Existence of equilibrium states.}\footnote{There are several ensembles in statistical mechanics: the microcanonical ensemble, where the number of particles and the energy are fixed, the canonical ensemble where the number of particles in the system is fixed while the energy fluctuates, and the grandcanonical ensemble where both the number of particles and the energy are allowed to fluctuate. Although different for finite systems, the three ensembles are usually equivalent in the thermodynamic limit.}
For $A\in \O^\infty$, consider the sequence of equilibrium
expectation values $\omega^{\Lambda_i}_\beta$ at inverse
temperature $\beta>0$. We assume existence of a limit of a
suitable (sub)sequence $\omega^{\Lambda_i}_\beta (\cdot ),$ as
$i\rightarrow\infty$. The limiting equilibrium state, $\omega^\R_\beta,$ is
$\a_\R^t$-invariant
\begin{equation}
\omega^\R_\beta(\a_\R^t (A))=\omega^\R_\beta (A) \; ,
\label{timetrans1}
\end{equation}
for $A\in \O^\R$ and $t\in {\mathbf R}$. Moreover, it satisfies
the Kubo-Martin-Schwinger (KMS) condition, which says that, for $A,B$ in a norm-dense subalgebra of $\O^\R$,
\begin{equation}
\omega^\R_\beta(A\a_\R^t (B))=\omega^\R_\beta (\a_\R^{t-i\beta}
(B) A) \; . \label{KMS2}
\end{equation}
\end{itemize}
The following principle concerning thermodynamic limits will be assumed henceforth; (but see for example [Ru1]).

\bigskip
\noindent {\bf Principle concerning thermodynamic limits}

\noindent {\it Let $I\subset {\mathbf R}$ be an interval of time, $E$ bounded subset of ${\mathbf R}^3,$ and $\epsilon >0.$ Then there exists a compact set $\Lambda(\epsilon, I, E)\subset {\mathbf R}^3$ , $|\Lambda (\epsilon, I, E)|<\infty,$ such that, $\forall \Lambda \supset \Lambda (\epsilon, I, E),$
\begin{equation*}
\omega^\Lambda_\beta (e^{itH^\Lambda}A e^{-itH^\Lambda}B) = \omega^\R_\beta (\a_\R^t (A)B)+ O(\epsilon),
\end{equation*}
$\forall A,B \in \O^E,$ $\forall t\in I.$}
\bigskip

For the sake of clarity of exposition, we will assume, throughout the following discussion, that reservoirs are finite and take the thermodynamic limit of suitable quantities at the end of every argument. We note, however, that one may work directly with reservoirs in the thermodynamic limit (see for example [BR,H,Sa]). KMS states satisfy certain stability properties which justify to view them as equilibrium states of thermal reservoirs; (see [HHW,HKTP] for a detailed discussion of this point). They give rise to an eigenvector of the Liouvillean, obtained via the GNS construction, corresponding to the simple eigenvalue 0.

\bigskip
\noindent{\bf Thermodynamic processes}

We first sketch what we mean by different thermodynamic processes before considering specific ones, later. Consider a thermodynamic system $\S$ coupled to $n$ reservoirs, $\R_1,\cdots ,\R_n.$ We assume that the reservoirs are finite and then take the thermodynamic limit of suitable quantities. The initial state, ${\mathsf P},$ of the coupled system $\S\vee(\bigvee_{i=1}^n \R_i),$ is normal relative to ${\mathsf P}^\S\otimes (\bigotimes_{i=1}^n {\mathsf P^{\R_i}}),$ where ${\mathsf P}^{\R_i}$ is an equilibrium state of $\R_i.$ The dynamics is generated by the Hamiltonian
\begin{equation*}
H(t)=H^\S(t)+\sum_{i=1}^n H^{\R_i} ,
\end{equation*}
where
\begin{equation}
\label{Hsigma}
H^\S(t)=H_0^\S (t)+ I^{\S\vee(\bigvee_{i=1}^n \R_i)}(t),
\end{equation}
and $I^{\S\vee(\bigvee_{i=1}^n \R_i)}(t)\in \O^\S\otimes(\bigotimes_{i=1}^n \O^{\R_i})$ describes interactions between $\S$ and the reservoirs. The {\it density matrix} ${\mathsf P}_t$ of the total system at time $t$ satisfies the Liouville equation
\begin{equation}
\dot{{\mathsf P}}_t = -i[H(t),{\mathsf P}_t] =-i \L(t)\P_t,
\end{equation}
and ${\mathsf P}_{t=0}={\mathsf P}.$ Note that one may work directly in the thermodynamic limit. On the state space determined by the initial state $\omega=\omega^\S\otimes \omega^{\R_1}\otimes\cdots\otimes\omega^{\R_n}$ via the GNS construction, the dynamics is unitarily implemented by a propagator $U(t,s),$ which is generated by time-dependent Liouvilleans $\L(t).$ A basic problem in quantum statistical mechanics
is to establish the existence of the thermodynamic limit of the
following quantities, [HHW, AWo,AWy,Rob,BR],
\begin{align}
\rho_t(\cdot) &:=\lim_{TD}Tr(\cp_t \cdot) \; , \; (true \; state \; of \; the \; composed \; system) \\
\rho_t^\S &:= \rho_t |_{\O^\S\otimes\id^\R} \; , \; (restriction \; to \; the \; subsystem \; \S),
\end{align}
and of the dynamics
$\alpha_t.$ Here, ``$\lim_{TD}$'' refers to the thermodynamic limit of the reservoirs. Moreover, for a thermodynamic system $\S$ coupled to a single reservoir $\R$, the {\it instantaneous equilibrium state} corresponding to the Hamiltonian $H(t)=H^\S(t)+H^\R$ at inverse temperature $\beta$ is given by
$${\mathsf P_t^\beta}:=\frac{e^{-\beta H(t)}}{Z_\beta(t)},$$
where $Z_\beta(t)=\tr e^{-\beta H(t)}.$ A standard problem is to establish the existence of the thermodynamic limit of instantaneous equilibrium states,
\begin{equation}
\omega_t^\beta (\cdot) = \lim_{TD}\tr ({\mathsf P}_t^\beta \cdot) .
\end{equation}
We refer the reader to [BR,Ru1] for a rigorous discussion of the existence of these limits for a large class of systems.

The choice of reservoirs $\R_1, \cdots, \R_n,$ the initial state ${\mathsf P}$ of $\S\vee(\bigvee_{i=1}^n \R_i),$
and the dynamics $\{H^\S(t)\}_{t\in {\mathbf R}},$ determine a trajectory of states $\{\rho_t^\S\}$ of $\S,$ where
\begin{equation}
\rho_t^\S (A) := \lim_{TD} \tr ({\mathsf P}_t A\otimes\unit),
\end{equation}
$A\in \O^\S .$

Isothermal processes correspond to diathermal contacts of $\S$ to
a single heat bath, $n=1$ (or, equivalently, several heat baths but with the {\it same}
temperature). Adiabatic processes are processes of an isolated system. Circular or cyclic
processes are processes with the property that $H^\S(t+t_*)=H^\S(t)$, for a period $t_*<\infty .$

The reservoirs considered in this paper are usually formed of ideal Bose-/ Fermi gases, such as black-body radiation, noninteracting magnons in a magnet, or electrons in a metal. Moreover, a typical thermodynamic system $\S$ may be an array of quantum spins, discrete quantum dots, or interacting bounded subsystems of a reservoir $\R.$ The mathematical methods used in our analysis are the algebraic formulation of quantum statistical mechanics [HHW, ArWo, ArWy], scattering theory as developed in [Rob, He], and spectral and resonance theory (spectral deformations, Mourre theory, Fermi's Golden Rule); see for example [A-S,BFS,DJ,DJP,FM1,2,FMUe,JP1,2,3,M1,2,MMS1,2]. We make the following assumption regarding the idealized models considered in this paper. 

\begin{itemize}

\item[(A)] The subsystem $\S$ has a finite-dimensional Hilbert space ($dim \H^\S=d<\infty$), the reservoirs $\R_i, i=1,\cdots,n,$ are formed of non-interacting bosons or fermions. Interactions between $\S$ and $\R_1,\cdots,\R_n$ are described by local operators affiliated with the kinematical algebra of the coupled system and with certain regularity properties; see [A-S,BFS,JP2] for concrete examples for which the analysis sketched in this paper is applicable.

\end{itemize}


\section{Internal- and heat energy, work, entropy and the $1^{st}$  Law}

As in the previous section, consider a finite system $\S$ ($dim (\H^\S)=d<\infty$) coupled diathermally to several reservoirs $\R_1,\cdots, \R_n.$ As mentioned earlier, the reservoirs are first assumed to be finite, and the thermodynamic limit of suitable quantities will be taken at the end of each argument. The {\it internal energy} of $\S$ is defined by
\begin{equation}
U^\S(t):=\rho_t(H^\S(t)) ,
\end{equation}
where $\rho_t$ is the true state of the total system, and $H^\S(t)$ is as in (\ref{Hsigma}) (Sect. 3).
The rate of change of {\it heat energy} is given by
\begin{align*}
\frac{\delta Q(t)}{dt}
&:=-\sum_{i=1}^n\frac{d}{dt}\rho(H^{\R_i})=-i\sum_{i=1}^n\rho_t([H(t),H^{\R_i}])
\\
&= i\sum_{i=1}^n \rho_t([H^{\R_i},I^{\S\vee(\bigvee_{i=1}^n \R_i)}(t)]) =:
\sum_{i=1}^n \frac{\delta Q^{\R_i}(t)}{dt} \; ,
\end{align*}
where $\delta (\cdot)$ denotes the {\it imperfect} or {\it inexact}
differential of $(\cdot)$. It follows that
\begin{equation}
\dot{U}^\S(t)-\frac{\delta
Q(t)}{dt}=\rho_t(\dot{H}^\S(t))=:\frac{\delta A(t)}{dt}.\label{FirstLaw}
\end{equation}
The thermodynamic limit for the reservoirs exists on both sides of this identity. Eq. (\ref{FirstLaw}) is nothing but the expression of the $1^{st}$ Law of Thermodynamics.

Next, we define the {\it relative entropy} of $\S,$ with respect to the {\it reference state} $${\mathsf P}^\R:=\frac{1}{d}{\mathbf 1}^\S\otimes_{i=1}^n{\mathsf P}^{\R_i},$$
as
\begin{align}
S^\S (t) &:=  -k_B \lim_{TD} Tr(\cp_t [ \log \cp_t - \log \cp^{\R} ] ) \\
         &= -k_B \lim_{TD} Tr(\cp_t [ \log \cp_t -\sum_{i=1}^n \log \cp^{\R_i}+\log d ] ) \\
         &= -k_B \lim_{TD} Tr(\cp_t [ \log \cp_t + \sum_{i=1}^n ( {\beta}_i H^{\R_i} + \log Z^{\R_i} )+\log d ]) \; ,
\end{align}
where $k_B$ is the Boltzmann constant, and $Z^{\R_i}=\tr (e^{-\beta_i H^{\R_i}}).$ Since we are assuming that $\H^\S$ is finite-dimensional, this quantity is well-defined, and the thermodynamic limit of the reservoirs can be taken. The usefulness of this notion of entropy will become apparent soon. 

A well-known trace inequality (see [BR]) says that
\begin{equation*}
Tr(B\log B) \ge Tr(B\log A)+Tr(B-A),
\end{equation*}
for $A$ and $B$ positive, and bounded operators. This inequality implies that the relative entropy of $\S$
has a definite sign, for all $t\in {\mathbf R},$\footnote{Another property of relative entropy is its {\it strong subadditivity} (see for example [LR]).}
\begin{equation}
S^\S (t)\le 0 \; .
\label{EntropySign}
\end{equation}

The quantities $Tr \cp_t\log \cp_t$ and $Tr \cp_t \log Z^{\R_i}$ are
time-independent.  Therefore, the rate of
change of entropy is
\begin{align}
\dot{S}^\S (t) &= -\sum_i \frac{1}{T_i}
\frac{d\rho_t(H^{\R_i})}{dt} \\
               &= \sum_i \frac{1}{T_i}\frac{\delta Q^{\R_i}(t)}{dt} \; .
\end{align}

Note that if the limit of the rate of entropy production
$${\cal E}:=-\lim_{t\rightarrow\infty}\dot{S^\S}(t)$$
exists then ${\cal E}\ge 0,$ as follows from the  upper bound in (\ref{EntropySign}). This bound on ${\mathcal E}$ is closely related to the Second Law of Thermodynamics, as we will see later.


\section{Isothermal processes, return to equilibrium, and the isothermal theorem}

In this section, we consider a system $\S$ diathermally coupled to a single heat bath $\R$ at temperature $T^{\R}>0.$ As shown in the previous section, 
\begin{equation}
\dot{U}^\S(t)=\frac{\delta}{dt} Q^\S(t) + \frac{\delta}{dt}A^\S (t) ,
\end{equation}
and
\begin{equation}
\dot{S}^\S (t)=\frac{1}{T^\R}\frac{\delta}{dt}Q^\S .\label{EntropyRate}
\end{equation}
We begin the non-trivial part of our analysis by considering an example of an irreversible thermodynamic process, that plays an important role in deriving the Zeroth Law of Thermodynamics.


\bigskip
\noindent{\bf Return to equilibrium} [JP1,2,BFS,M1,2,FM1,A-S]

\noindent {\it If $(\S,\R)$ belongs to the class of model systems satisfying Assumption (A), Sect. 3, with the properties that
\begin{itemize}
\item[(i)]
$$\int^\infty dt \|(H^\S(t)-H_\infty^\S)((H_\infty^\S+i)^{-1})\|<\infty,$$ 
with {\it form factors} in the interaction Hamiltonian that are sufficiently regular in the infrared, ie, for small wave vectors (see, for example, [BFS,FM1] for precise statements), and
\item[(ii)] Fermi's Golden rule (when $\S$ is coupled to the reservoirs) holds, in the sense that, to second order in perturbation theory, all the eigenvalues of the unperturbed Liouvillean, except a single one at 0, become resonances when the perturbation is turned on (ie, develop an imaginary part),
\end{itemize}
then
\begin{equation}
\rho_t\rightarrow \omega^\beta ,
\end{equation}
where $\omega^\beta$ is the equilibrium state of the coupled system at inverse temperature $\beta$ corresponding to the Hamiltonian $H_\infty=H_\infty^\S+H^\R.$}
\bigskip

Here the problem of proving the property of return to equilibrium is viewed as a spectral problem. The coupled system exhibits return to equilibrium if 0 is a simple eigenvalue of the (standard) Liouvillean corresponding to $H_\infty,$ and the spectrum of the Liouvillean away from zero is continuous; (see for example [JP1,2,BFS]). Using different methods of spectral theory, the property of
return to equilibrium has been established for a variety of
quantum mechanical systems: complex spectral deformation techniques for the
spin-boson model [JP1,JP2], Feshbach map and operator-theoretic renormalization group methods, in conjunction with complex spectral deformations, for a small
system coupled to a thermal reservoir of photons [BFS], and an
extension of Mourre's positive commutator method, together with a
Virial Theorem, for a small system coupled to a thermal
bath of free bosons, [M1,FM1]. The general formalism used in these papers is based on important insights in [HHW]. Positive commutator methods used in studying return to equilibrium
have been extended to studying {\it thermal ionization} of atoms and
molecules coupled to the radiation field in [FM2,FMS], and to prove the property of return to equilibrium for a variety of further, physically interesting systems, e.g., an impurity spin coupled to a bath of noninteracting magnons in a magnet, or a quantum dot coupled to nonrelativistic electrons in a metal; (see [A-S]).


Suppose that the coupled system $\S\vee\R$ has the property of return to
equilibrium. What happens if the coupling is {\it slowly} turned off, after
return to equilibrium, e.g., by {\it quasi-statically}
removing the contact between $\R$ and $\S$? What characterizes {\it reversible} isothermal processes? The answer to these questions relies on the so called {\it isothermal theorem}, which is an adiabatic theorem for states close to thermal equilibrium states.

We consider a system $\S\vee\R$ directly in the thermodynamic
limit, and study a process with Liouvillean $\L^\tau(t),$ given by $\L^\tau(t)\equiv \L(s)$,
where the rescaled time is $s:=\frac{t}{\tau},$ and $\{ \L(s)\}$ is a family of time-dependent ``standard'' Liouvilleans. We assume that the operators $\L(s)$
have a common dense domain of definition, for all $s\in I$, where $I\subset
{\mathbf R}$ is a compact interval. Moreover, we assume that, for all
$s\in I$, $(\L(s)+i)^{-1}$ is differentiable in $s$,
$\L(s)\frac{d}{ds}(\L(s)+i)^{-1}$ is uniformly bounded,
$\sigma_{pp}(\L(s))=\{ 0 \}$ and $\sigma(\L(s))\backslash \{0\} =
\sigma_c(\L(s))$, and that the projection, $P(s),$ onto the eigenstate
corresponding to the eigenvalue 0 of $\L(s)$ is twice differentiable in $s,$ for almost all $s\in I$. Note that
$P(s)$ projects onto the instantaneous equilibrium state, or
reference state, $\omega^\beta_{\tau s}$ at time $t=\tau s$. We
are interested in the {\it quasi-static limit}, $\tau\rightarrow\infty.$
Physically, this limit corresponds to $\tau\gg\tau_R$, where
$\tau_R=\max_{s\in I}\tau_R (s),$ and $\tau_R(s)$ is the {\it relaxation time} to equilibrium of $\L(s),$ $s\in I.$

\bigskip
\noindent{\bf Isothermal Theorem} [A-SF1]

\noindent {\it Under the hypotheses described above,
$$\rho_{\tau s}(A)=\omega^\beta_{\tau s}(A) + o(1),$$
as $\tau\rightarrow\infty$, $\forall A\in\O^\S\otimes\O^\R,$ and
$\forall s\in I$, where $I \subset{\mathbf R}$ is an arbitrary
compact interval, (ie, $\rho_{\tau s}(A)-\omega_{\tau s}^\beta (A)$ tends to 0, as $\tau\rightarrow\infty$).}
\bigskip

The proof of this theorem can be found in [A-SF1]. It relies on a slight generalization of results in [AE,Te]. To get a quantitative estimate on the rate of convergence to the
quasi-static limit, we need more precise information on the spectrum
of the standard Liouvilleans $\L(s)$. The hypotheses of the isothermal theorem can be verified for classes of explicit quantum
mechanical systems, including ones for which the property of return to equilibrium has been established, [A-S].

Next, we sketch several consequences of this theorem,
clarifying the notions of heat energy and reversibility in isothermal
processes and emphasizing the usefulness of relative
entropy. Without loss of generality, we first treat $\R$ as a finite system, before taking the thermodynamic limit of suitable quantities.

Consider an isothermal process of $\S\vee\R$ from time $t_0=\tau s_0$
until some time $t_1=\tau s_1$, for $s_0$ and $s_1$ fixed, in the quasi-static limit where
$\tau\rightarrow\infty$, and for an initial state
$\rho=\omega^\beta.$ Then we have the following results.

\begin{itemize}

\item[(i)] {\it Reversible isothermal processes are the same as
``quasi-static'' isothermal processes ($\tau \gg \tau_R$).} This just means that the {\it true} state of the system coincides with the {\it instantaneous equilibrium state}, asymptotically when $\tau\rightarrow\infty .$ 
\bigskip

For {\it reversible} isothermal processes, the entropy of a system $\S$ coupled to a heat bath $\R$ is defined by 
\begin{align}
S^\S_{rev}(t) &=-\lim_{TD}k_B Tr (\cp_t^\beta[\log \cp_t^\beta - \log \cp^\R]) \\
              &= \lim_{TD}[k_B \omega_t^\beta (\beta H^\S (t)) +k_B \log \frac{Z_\beta (t)}{Z^\R}-k_B\log d] \\
              &= \lim_{TD}[\frac{1}{T^\R}(U^\S_{rev}(t)-F^\S(t))]-k_B\log d \; ,
\end{align}
where $F^\S(t)=-k_B \log \frac{Z_\beta (t)}{Z^\R}$ is the
{\it free energy} of $\S,$ and $\P_t^\beta$ denotes the instantaneous equilibrium state of $\S\vee\R$ at time $t.$ Using
the isothermal theorem, one may replace $\omega_{\tau s}^\beta$ by the true state 
$\rho_{\tau s}$ of $\S\vee\R,$ up to an error that vanishes in the
quasi-static limit: Hence, in the thermodynamic limit,

\begin{eqnarray}
T^\R\Delta S^\S_{rev} &=& \omega_{\tau s_1}^\beta (H^\S(s_1)) - \omega_{\tau s_0}^\beta (H^\S(s_0)) - \int_{s_0}^{s_1}ds \omega_{\tau s}^{\beta}(\dot{H}^\S(s)) \nonumber \\
                   &=& \Delta U^\S -\Delta A + o(1) \\
                   &=& \Delta Q + o(1) \; ,\label{1Law}
\end{eqnarray}
with 
$$\Delta A=F^\S(\tau s_1)-F^\S(\tau s_0)+ o(1).$$
Here, we have made use of the isothermal theorem in the second step and
the $1^{st}$ Law of Thermodynamics in (\ref{1Law}). We have just
sketched the proof of the following claim, which asserts the
equivalence of the definition of entropy in equilibrium
statistical mechanics and relative entropy in non-equilibrium
quantum statistical mechanics, in the quasi-static limit.

Clearly, (\ref{1Law}) and (\ref{EntropyRate}) imply that

\bigskip
\item[(ii)]{\it $\Delta S_{rev}^\S = \Delta S^\S + o(1),$ as
$\tau\rightarrow\infty$.}
\bigskip

Furthermore, if one slowly removes the contact between $\R$ and
$\S$ the state of $\S$ approaches a {\it Gibbs} state at inverse
temperature $\beta_\R$, independently of the nature of the diathermal contact. {\it This is part of the $0^{th}$ Law of Thermodynamics}.

\item[(iii)]
{\it If $H^\S(t)\rightarrow H^\S_\infty \in \O^\S$ then $\rho_{\tau s}$
tends to the {\it Gibbs} state for $H^\S_\infty$ at inverse temperature $\beta_\R,$ as
$\tau\rightarrow\infty$ and $s\rightarrow\infty$; (for a more precise formulation of this result, see [A-SF4]).}

\end{itemize}


\section{Clausius' and Carnot's formulations of the $2^{nd}$ Law}

We consider a thermodynamic system $\S$ coupled diathermally to heat baths $\R_1,
\cdots, \R_n$, with $n\ge 2$. We have shown that, for diathermal
contacts,
\begin{equation}
 -\infty< S^\S (t) \le 0 \; ,
\end{equation}
and
\begin{equation}
\dot{S}^\S (t)=\sum_i \frac{1}{T_i}\frac{\delta Q^{\R_i}(t)}{dt}
\; .
\end{equation}

Suppose that at least two reservoirs are at different temperatures. Under certain conditions on the coupling, and for the class of idealized model systems discussed above (see Assumption (A), Sect. 3), one can show that the state of the coupled system, $\rho_t,$ converges to a non-equilibrium steady state (NESS),
$$\rho^{NESS}:=w^*-\lim_{t\rightarrow\infty}\rho_0 \circ
\alpha_t,$$ 
(or, more generally, $w^*-\lim_{T\rightarrow\infty}\frac{1}{T}\int_0^{T}\rho_0 \circ \alpha_t dt$). This has been proven recently in several examples using
different approaches: In [FMUe,Ru2,3] algebraic scattering theory is used, and 
one has to establish the existence of scattering
endomorphisms. The results are based on work of [He,Rob,BoMa]. As an alternative approach in [JP3,MMS1,2], a NESS is related to a zero-energy resonance of
the adjoint of the so called C-Liouvillean. In this setting, 
one can prove an adiabatic theorem for states close to non-equilibrium
steady states; see[A-S].

\bigskip 
\noindent {\bf Clausius' formulation of the $2^{nd}$ Law}

\noindent {\bf Theorem.}

{\it Assume that
$$H^\S(t)\rightarrow H^\S_\infty\in \O^\S\otimes\O^\R,$$
as $t\rightarrow\infty$. If
$$\rho_t\rightarrow_{t\rightarrow\infty}\rho^{NESS},$$ then
\begin{align}
& (i) \sum_{i=1}^n \frac{\delta Q^{\R_i}}{dt} \rightarrow 0 \\
& (ii) \dot{S}^\S (t) \rightarrow - {\cal E} \le 0 \\
& (iii)\lim_{t\rightarrow\infty}\sum_{i=1}^n\frac{1}{T_i}\frac{\delta
Q^{\R_i}}{dt} =-{\cal E} \le 0 \; ,
\end{align}
where ${\cal E}$ is the entropy production rate.}
\bigskip

Clausius' formulation of the Second Law of Thermodynamics is a straightforward corollary of this theorem. We consider two resevoirs, $\R_1$ and $\R_2,$ with $T_1> T_2.$ Denote by ${\mathcal P}^\R(t):=\frac{\delta}{dt}Q^\R (t),$ the heat current out of reservoir $\R.$ It follows from $(i)$ that
\begin{equation}
\lim_{t\rightarrow\infty}{\mathcal P}^{\R_1}(t)+{\mathcal
P}^{\R_2}(t)=0 ,
\end{equation}
and from $(iii)$ that
\begin{equation}
\lim_{t\rightarrow\infty}(\frac{1}{T_1}-\frac{1}{T_2}){\mathcal P}^{\R_1}(t)\le 0 .
\end{equation}
Since $T_1>T_2,$ it follows that ${\mathcal P}^{\R_1}=\lim_{t\rightarrow\infty}{\mathcal P}^{\R_1}(t)\ge 0,$ ie, heat flows from the hot reservoir to the cold one. For small enough coupling, one can usually show {\it strict positivity of the entropy production rate}, ${\cal E,}$ by computing ${\cal E}$ perturbatively; (see [FMUe, JP3, MMS1,2]). A study of transport phenomena between two reservoirs formed of free fermions at different
temperatures/chemical potentials and coupled through bounded local
interactions has been presented in [FMUe]. After showing
convergence of the true state of the coupled system to a NESS by using scattering theory, and establishing strict positivity of the entropy production rate,
these authors show that the {\it Onsager reciprocity relations} and Ohm's Law hold to first non-trivial order in the coupling constant, for contacts allowing exchange of particles between the reservoirs. Furthermore, in [JOP1,2] linear response theory is studied from the point of view of the algebraic formulation of quantum statistical mechanics, and the Green-Kubo formula and Onsager reciprocity relations for heat fluxes generated by temperature gradients are established.


Next, we discuss Carnot's formulation of the Second Law of Thermodynamics. For the class of models discussed above, we consider a cyclic thermodynamic process, with $H^\S (t+\tau_*)=H^\S (t)$, for some period $\tau_*<\infty .$ For $t\in [0,\tau_*),$ let 
$$\omega_t^{per}:=\lim_{N\rightarrow\infty}\rho_{t+N\tau_*},$$
which is a time-periodic state with period $\tau_*.$ For some class of model systems, with small enough coupling, one can show that, after very many periods, the state of the coupled system approaches $\omega^{per};$ see [A-SF2] (and also [FMSUe]) for precise formulations and proofs.   

\bigskip
\noindent{\bf Cyclic thermodynamic processes and time-periodic states}

\noindent {\it For the class of models considered in Assumption (A) of Sect. 3, if $H^\S (t+\tau_*)=H^\S (t),$ for some period $\tau_*<\infty ,$ and the interaction Hamiltonian $\|I^{\S\vee(\bigvee_{i=1}^n\R_i)}\|$ is sufficiently regular and satisfies a Fermi Golden Rule condition then
\begin{equation*}
\rho_{t+N\tau_*} \rightarrow_{N\rightarrow\infty} \omega_t^{per} ,
\end{equation*}
for $t\in [0,\tau_*).$}
\bigskip

This is proven in [A-SF2] for fermionic reservoirs, by introducing the so called {\it Floquet Liouvillean} and relating the time-periodic state to a zero-energy resonance of the latter.\footnote{One can also prove this result for fermionic reservoirs using scattering theory and a norm-convergent Dyson-Schwinger series, as in [FMSUe,FMUe].}The existence of the limit and the absolute upper bound on the relative entropy $S^\S (t)$ imply that the {\it entropy production per cycle},
\begin{equation}
\Delta {\cal E} := -  \lim_{N\rightarrow\infty} \int_0^{\tau_*} dt \dot{S}^\S (t+N\tau_*),
\end{equation}
is {\it non-negative}.
Furthermore, for specific models, such as the one considered in [A-SF2], one can actually prove {\it strict positivity of entropy production per cycle}, which can be computed perturbatively, for small enough coupling. 

We now discuss {\it Carnot's formulation} of the Second Law of Thermodynamics. Suppose $\S$ is coupled to two reservoirs $\R_1$ and $\R_2,$ with $T_1>T_2.$ For a thermodynamic quantity $f,$ we set 
\begin{equation*}
\Delta f := \lim_{N\rightarrow\infty}[f((N+1)\tau_*) - f(N\tau_*)] ,
\end{equation*}
which is the change of $f$ in one cycle, after very many periods. Suppose the state of the coupled system converges to $\omega^{per},$ after very many periods. Since $H^\S(t+\tau_*)=H^\S(t),$ it follows that 
\begin{equation}
\Delta U^\S =0 \; .
\end{equation}
Furthermore, from the fact that $\Delta {\cal E}\ge 0,$ it follows that
\begin{equation}
\frac{\Delta Q^{\R_1}}{T_1}+\frac{\Delta Q^{\R_2}}{T_2} = -\Delta
{\cal E} \le 0 \; . \label{2Law}
\end{equation}
Suppose that the system $\S$ is a heat engine, ie, performs work during each period, 
\begin{equation}
\Delta A^\S = \Delta Q^{\R_1}+\Delta Q^{\R_2} \ge 0.
\end{equation}


The fact that $T_1 \ge T_2$ and the arguments used in the proof of {\it Clausius'} formulation imply that $\Delta Q^{\R_1} \ge 0.$ The following result yields
 Carnot's formulation of the $2^{nd}$ Law of Thermodynamics.

\bigskip
\noindent{\bf Carnot's formulation of the $2^{nd}$ Law}

\noindent{\it Assume that $T_1\ge T_2$. Then
\begin{eqnarray}
0 \le \eta^\S &:=& \frac{\Delta A}{\Delta Q^{\R_1}} = 1 + \frac{\Delta Q^{\R_2}}{\Delta Q^{\R_1}}\\
              & \le & 1 - \frac{T_2}{T_1} := \eta^{Carnot} \; .
\end{eqnarray}}
\bigskip

It is important to note that this result follows from
the absolute upper bound on relative entropy and the existence of time periodic
states in the large-time limit, without any further assumptions. The difference
$\eta^{Carnot}-\eta^\S$ can be computed explicitly in terms of the
entropy production per cycle, [A-SF2], which is a quantity that can be computed perturbatively; see [FMUe]. Inequality (\ref{2Law}) can easily be generalized to 
\begin{equation*}
\sum_{i=1}^n \frac{\Delta Q^{\R_i}}{T_i}\le 0,
\end{equation*}
for an arbitrary number, $n<\infty,$ of reservoirs. This can be used to prove that a certain notion of entropy increases in adiabatic processes.

\bigskip
\noindent{\bf Acknowledgements}

\noindent We thank Gian Michele Graf, Marco Merkli and Daniel Ueltschi for many helpful discussions.



\begin{thebibliography}{9}

\bibitem[A-S]{} Abou-Salem, W., {\it Nonequilibrium quantum statistical mechanics and thermodynamics}, ETH-Diss. 16187 (2005)

\bibitem[A-SF1]{} Abou-Salem,W., Fr\"ohlich, J., {\it Adiabatic theorems and reversible isothermal
processes}, Lett. Math. Phys. {\bf 72}, 153-163 (2005)


\bibitem[A-SF2]{} Abou-Salem,W., Fr\"ohlich, J.,{\it Cyclic thermodynamic processes and entropy production}, to appear in J. Stat. Phys.

\bibitem[A-SF3]{} Abou-Salem,W., Fr\"ohlich, J., {\it Adiabatic theorems for quantum
resonances}, to appear in Commun. Math. Phys.

\bibitem[A-SF4]{} Abou-Salem, W., Fr\"ohlich, J., in preparation

\bibitem[ArWo]{} Araki, H., Woods, E., {\it Representations of the canonical
    commutation relations describing a non-relativistic infinite free bose
    gas}. J. Math. Phys. {\bf 4}, 637-662 (1963)

\bibitem[ArWy]{} Araki, H., Wyss, W., {\it Representations of canonical anticommutation
relations}, Helv. Phys. Acta {\bf 37}, 136 (1964)

\bibitem[AE]{} Avron,J.E., Elgart, A., {\em Adiabatic theorem
without a gap condition}, Commun. Math. Phys. {\bf 203},
445-463 (1999)


\bibitem[BFS]{} Bach, V., Fr\"ohlich, J., Sigal, I.M., {\it Return to
    Equilibrium}. J. Math. Phys. {\bf 41 no 6}, 3985-4061 (2000)

\bibitem[BLR]{} Benetto, F., Lebowitz, J.L., Rey-Bellet, L., {\it  Fourier's Law: A Challenge to Theorists}, Mathematical Physics 2000, 128-150, Imperial College Press, London, 2000

\bibitem[BoMa]{} Botvich, D.D., Malyshev, V.A., {\it Unitary equivalence of temperature dynamics for ideal and locally perturbed Fermi gas}, Commun. Math. Phys. {\bf 61}, 209 (1978)

\bibitem[Boy]{} Boyling, J.B., {\it An axiomatic approach to classical thermodynamics}, Proc. R. Soc. London A {\bf 329}, 35-70 (1972)

\bibitem[BR]{} Bratteli, O., Robinson, D., {\it Operator Algebras and
    Quantum Statistical Mechanics 1,2}, Texts and Monographs in Physics,
    Springer-Verlag Berlin, 1987

\bibitem[DJ]{} Derezi\'nski, J., Jaksi\'c, V., {\it Spectral theory of
    Pauli-Fierz operators},  Ann. Henri Poincare {\bf 4}, 739 (2003)

\bibitem[DJP]{}Derezi\'nski, J., Jaksi\'c, V.,Pillet, C.-A., {\it Perturbation theory of KMS
states},  Rev. Math. Phys. {\bf 15}, 447 (2003)

\bibitem[FM1]{} Fr\"ohlich, J., Merkli, M.,
{\em Another return of ``return to equilibrium''}, Commun. Math.
Phys. {\bf 251}, 235-262 (2004)

\bibitem[FM2]{} Fr\"ohlich, J., Merkli, M., {\it Thermal Ionization.} Math. Phys. Analysis and Geometry {\bf 7}, 239-287 (2004)

\bibitem[FMS]{}Fr\"ohlich, J., Merkli, M., Sigal, I.M., {\it Ionization of atoms
in a thermal field.}, J. Stat. Phys. {\bf 116}, 311-359 (2004)

\bibitem[FMUe]{} Fr\"ohlich, J., Merkli, M. and Ueltschi, D., {\em Dissipative
transport: thermal contacts and tunnelling junctions}, Ann. Henri
Poincar\'e {\bf 4}, 897-945 (2003)

\bibitem[FMSUe]{}Fr\"ohlich, J., Merkli, M.,  Schwarz, S., and Ueltschi, D.,
 {\it Statistical mechanics of thermodynamic processes},
in {\it A garden of quanta}, 345-363, World Sci. Publishing, River
Edge, New Jersey, 2003

\bibitem[H]{} Haag, R., {\it Local Quantum Physics. Fields, Particles,
    Algebras}. Text and Monographs in Physics. Springer-Verlag Berlin (1992)

\bibitem[HHW]{}Haag, R., Hugenholtz, N. M., Winnink, M., {\it On the equilibrium
states in quantum statistical mechanics.}, Commun. Math. Phys.
{\bf 5}, 215--236 (1967)

\bibitem[HKTP]{} Haag, R., Kastler, D., Trych-Pohlmeyer, E., {\it Stability and equilibrium states}, Commun. Math. Phys. {\bf 38}, 213 (1974)

\bibitem[He]{} Hepp, K., {\it Rigorous results on the $s-d$ model of the Kondo
effect.} Solid State Commun. {\bf 8} , 2087--2090 (1970)



\bibitem[JP1]{} Jaksi\'c, V., Pillet, C.A., {\it On a Model for Quantum
    Friction II. Fermi's Golden Rule and Dynamics at Positive
    Temperature}, Commun. Math. Phys. {\bf 176}, 619-644 (1996)

\bibitem[JP2]{} Jaksi\'c, V., Pillet, C.A., {\it On a Model for Quantum
    Friction III. Ergodic Properties of the Spin-Boson
    System}, Commun. Math. Phys.  {\bf 178}, 627-651 (1996)

\bibitem[JP3]{} Jaksi\'c, V., Pillet, C.-A.:{\em Non-equilibrium
steady states of finite quantum systems coupled to thermal
reservoirs}, Commun. Math. Phys. {\bf 226}, 131-162 (2002)

\bibitem[JOP1]{} Jaksic, V., Ogata, Y., Pillet, C.-A., {\it The Green-Kubo formula and the Onsager reciprocity relations in quantum statistical mechanics}, [Texas mp-arc 2005 preprint]

\bibitem[JOP2]{} Jaksic, V., Ogata, Y., Pillet, C.-A., {\it Linear response theory in quantum statistical mechanics}, [Texas mp-arc 2005 preprint]


\bibitem[LR]{} Lieb, E., Ruskai, M.-B, {\it Proof of the strong subadditivity of quantum-mechanical
entropy}, J. Math. Phys. {\bf 14}, 1938-1941 (1973)


\bibitem[LY]{} Lieb, E., Yngvason, J., {\it The mathematical structure of the second law of thermodynamics}, in {\it Current Developments in Mathematics, 2001}, International Press, Cambridge, 2002

\bibitem[M1]{} Merkli, M., {\it Positive Commutator Method in Non-Equilibrium Statistical Mechanics}, Commun. Math. Phys. {\bf 223}, 327-362 (2001)

\bibitem[M2]{} Merkli, M.,{\it Stability of equilibria with a condensate}, Commun. Math. Phys. {\bf 257}, 621-640 (2005)

\bibitem[MMS1]{} Merkli, M., M\"uck, M. and Sigal, I.M., {\it Instability of equilibrium states for coupled heat reservoirs at different temperatures}, [axiv:math-ph/0508005]

\bibitem[MMS2]{} Merkli, M., M\"uck, M. and Sigal, I.M., {\it Theory of nonequilibrium stationary states as a theory of resonances. Existence and properties of NESS}, [arxiv:math-ph/0603006]


\bibitem[Rob]{} D. Robinson, {\it Return to Equilibrium}, Commun. Math. Phys. {\bf
31}, 171--189 (1973)


\bibitem[Ru1]{}Ruelle, D., {\it Statistical Mechanics. Rigorous results}, Reprint
of the 1989 edition. World Scientific Publishing Co., Inc., River
Edge, NJ; Imperial College Press, London, 1999


\bibitem[Ru2]{} Ruelle, D., {\it Entropy production in quantum spin systems},
Comm. Math. Phys. {\bf 224} , no. 1, 3-16 (2001)

\bibitem[Ru3]{} Ruelle, D. {\it Natural nonequilibrium states in quantum
statistical mechanics},  J. Stat. Phys. {\bf 98} , no. 1-2, 57-75
(2000)


\bibitem[Sa]{} Sakai, S., {\it $C^*$-Algebras and $W^*$-Algebras}, Berlin, Springer, 1971

\bibitem[Te]{}Teufel, S., {\em A note on the adiabatic theorem},
Lett. Math. Phys. {\bf 58}, 261-266 (2001)










\end{thebibliography}
\end{document}